\documentclass[apj]{emulateapj}
\lefthead{TSUJIMOTO et al. }
\righthead{CHEMICAL EVOLUTION OF $^{244}$PU}

\def\ltsima{$\; \buildrel < \over \sim\;$}
\def\ltsim{\lower.5ex\hbox{\ltsima}}
\def\gtsima{$\; \buildrel > \over\sim \;$}
\def\gtsim{\lower.5ex\hbox{\gtsima}}
\def\ms{$M_{\odot}$ }
\def\msp{$M_{\odot}$}

\slugcomment{Accepted for publication in ApJ Letters}

\begin{document}
\title{Chemical evolution of $^{244}$Pu in the solar vicinity and its implication for the properties of r-process production}

\author{Takuji Tsujimoto\altaffilmark{1}, Tetsuya Yokoyama\altaffilmark{2}, and Kenji Bekki\altaffilmark{3}
}

\affil{$^1$National Astronomical Observatory of Japan, Mitaka-shi, Tokyo 181-8588, Japan; taku.tsujimoto@nao.ac.jp \\
$^2$Department of Earth and Planetary Sciences, Tokyo Institute of Technology, 2-12-1 Ookayama, Meguro-ku, Tokyo 152-8551, Japan \\
$^3$ICRAR, M468, The University of Western Australia, 35 Stirling Highway, Crawley Western Australia 6009, Australia
}

\begin{abstract}
Meteoritic abundances of r-process elements are analyzed to deduce the history of chemical enrichment by r-process from the beginning of disk formation to the present time in the solar vicinity, by combining the abundance information from  short-lived radioactive nuclei such as $^{244}$Pu with that from stable r-process nuclei such as Eu. These two types of nuclei can be associated with one r-process event and cumulation of events till formation of the solar system, respectively. With help of the observed local star formation history, we deduce the chemical evolution of  $^{244}$Pu and obtain three main results: (i) the last r-process event occurred 130-140 Myr before formation of the solar system, (ii) the present-day low $^{244}$Pu abundance as measured in deep sea reservoirs results from the low recent star formation rate compared to $\sim$4.5$-$5 Gyr ago, and (iii) there were $\sim$15 r-process events in the solar vicinity from formation of the Galaxy  to the time of solar system formation and $\sim$30 r-process events to the present time. Then, adopting a  reasonable hypothesis that a neutron star merger is the r-process production site, we find that the ejected r-process elements are extensively spread out and mixed with interstellar matter with a mass of $\sim 3.5\times10^6$\msp, which is about 100 times larger than that for supernova ejecta. In addition, the event frequency of r-process production is estimated to be one per about 1400 core-collapse supernovae, which is identical to the frequency of neutron star mergers estimated from the analysis of stellar abundances.
\end{abstract}


\section{Introduction}

The origin and evolution of r-process elements is one of the hottest research areas. The detection of a near-infrared light bump in the afterglow of a short-duration $\gamma$-ray burst \citep[e.g.,][]{Tanvir_13} has initiated the current understanding that a neutron star (NS) merger is a promising major site for r-process \citep[e.g.,][]{Lattimer_74, Eichler_89}. Two vigorous and successful research activities in particular have supported this understanding; one is r-process nucleosynthesis calculations in NS mergers \citep[e.g.,][]{Wanajo_14, Rosswog_14, Goriely_15} and the other is modeling Galactic chemical evolution of r-process elements in terms of the enrichment taking place through NS mergers \citep[e.g.,][]{Matteucci_14, Tsujimoto_14a, Wehmeyer_15, Ishimaru_15, Komiya_16}. 

In addition to the abovementioned studies, the features of observed stellar abundances in nearby dwarf galaxies present more direct evidences for NS mergers as the site for r-process. The properties characterizing NS merger events are their rarity and high yield of r-process elements per event. \citet{Tsujimoto_14a} have found that faint classical dwarf spheroidal galaxies (dSphs) exhibit a constant [Eu/H] of $\sim -1.3$ over the metallicity range of $-2$\ltsim [Fe/H]\ltsim$-1$, which implies  no Eu production events while more than $10^3$ core-collapse supernovae (CCSNe) increase the galactic Fe abundance. If CCSNe are associated with r-process production, a gradual increase in [Eu/H] by $\sim$ 0.35 to 1 dex is predicted for this [Fe/H] range, depending on the theoretical Eu yield from a CCSN \citep{Rauscher_02, Wanajo_13}. Moreover, \citet{Ji_16} have discovered in an ultra-faint dwarf galaxy, Reticulum II, a feature suggesting that a single event with an r-process yield as high as the amount expected from an NS merger remarkably enhances r-process abundances of this galaxy \citep[see also][]{Roederer_16}.    

Apart from capture of electromagnetic waves, we can assess the r-process site and properties by analysis of meteoritic abundances of r-process elements. In particular, abundances of short-lived radioactive nuclei, such as $^{129}{\rm I}$ or $^{244}{\rm Pu}$, produced by the r-process inside meteorites have been known to be used as a clock to measure the time interval, $t_{\rm LE}$, between the last r-process production event and solar system formation since the 1960's \citep{Reynolds_60, Wasserburg_69, Podosek_70}. From the timescale thus obtained, we can infer the rarity of r-process events. According to many reports, a relatively long timescale of $t_{\rm LE}\approx$100 Myr is deduced \citep[e.g.,][see \S 2.1]{Clayton_83, Dauphas_05, Lugaro_14}. On the other hand, meteoritic abundances of  radionuclides such as $^{26}{\rm Al}$ (half-life:~1.03 Myr) and $^{60}{\rm Fe}$ (half-life:~2.2 Myr) imply the injection from a nearby CCSN since their abundances are higher than the steady-state abundance inferred from $\gamma$-ray observations and Galactic chemical evolution \citep{Huss_09}. Thus, we anticipate that r-process production events may occur much less frequently than CCSNe.

A recent finding of the current low abundance of $^{244}{\rm Pu}$ in deep sea reservoirs highlights the rarity of r-process events \citep{Wallner_15}. It was found that the current abundance is lower than expected from continuous production in CCSNe by about two orders of magnitude. This current abundance together with that in the early solar system (ESS) indeed gives a piece of evidence for the chemical evolution of $^{244}{\rm Pu}$ in the solar vicinity, which is found to support the rarity of r-process events that is compatible with an NS merger frequency \citep{Hotokezaka_15}. Besides, $^{244}{\rm Pu}$ abundance could trace the local star formation history with a rigid time-resolution since its short half-life of 81 Myr should make its abundance sensitive to the star formation rate (SFR) in individual epochs. On the other hand, the local SFR($t$) has been well studied by several methods of age-dating of stars \citep{Hernandez_00, Rocha_00, Vergely_02, Cignoni_06, Fuchs_09}. These studies have established our current understanding that the local SFR($t$) is not constant: it is oscillatory \citep{Hernandez_00} or has a few bursting star formation events \citep{Rocha_00} that will significantly impact $^{244}{\rm Pu}$ time evolution.

A new angle to analyze meteoritic abundances of r-nuclides provides key information on the history of r-process enrichment. While abundances of short-lived radioactive r-nuclides essentially reflect only the last r-process event, abundances of stable r-nuclides reflect the cumulation of all r-process events till the ESS. Thus, combining the two pieces of information will lead to counting of the total number of events that contribute to the local r-process enrichment. In this {\it Letter}, we present $^{244}{\rm Pu}$ time evolution by combining analysis results of meteoritic abundances (\S 2) with the observed local star formation history (\S 3). Our study further explores the deduction of the properties of r-process production, i.e., its frequency and its propagation, both of which are of great significance but remain unsolved issues (\S 4). 

\section{Assessment from unstable and stable nuclei}

We start by extracting the information on local r-process enrichment in the ESS from analysis of meteoritic abundances. We estimate the following three parameters; (i) how much a single r-process event enriches the fraction of a particular r-nuclide in the surrounding interstellar matter (ISM) per volume, (ii) how many times r-process events occur locally and enrich the ISM from the initial star formation till the ESS, and (iii) when does the last r-process event occur before the ESS. For this purpose, we need two types of r-nuclides: short-lived radioactive r-nuclide and stable r-nuclide. In our analysis, we adopt $^{244}{\rm Pu}$ (half-life: 81 Myr), which is present in the form of fissiogenic xenon isotopes, as the former and Eu as the latter.

\subsection{Mass fraction of $^{244}$Pu in the ISM by a single r-process event}

Meteoritic abundances of short-lived radioactive r-nuclides are dominated by the nuclides ejected from the last r-process production event as long as the time interval between r-process events is much longer than half-lives of the r-nuclides. $^{244}{\rm Pu}$ along with other r-nuclides such as $^{247}{\rm Cm}$ and $^{129}{\rm I}$ meets this condition. From this fact, we will deduce how much a single r-process event enriches the surrounding ISM with r-nuclides.

First, we deduce the mass fraction of $^{244}{\rm Pu}$ at the ESS. The number fraction of $^{235}$U relative to hydrogen at the ESS is estimated to be $6.37\times 10^{-13}$ from meteorites, considering the radioactive decay (half-life: 4.47 Gyr) till the present over 4.56 Gyr \citep{Lodders_09}. Then from the abundance of $^{244}$Pu relative to $^{235}$U (0.008) at the ESS \citep{Turner_04, Turner_07} and a typical number density of hydrogen in the ISM (1 cm$^{-3}$), the number density of $^{244}$Pu at the ESS is deduced to be $5.10\times 10^{-15}$ cm$^{-3}$ with the corresponding mass fraction of $8.90\times 10^{-13}$. 

The amount of $^{244}$Pu thus obtained includes small contributions from a few events before the last one. By removing these small contributions, we obtain how much a single r-process event polluted the ISM with $^{244}$Pu at the ESS. For this estimate, two timescales are required. One is the time interval $t_{\rm LE}$ between the last r-process event and solar system formation. The other is the mean time interval $t_{\rm int}$ between individual r-process events before the ESS epoch. $t_{\rm LE}$ is necessary to know the remaining fraction of $^{244}$Pu owing to a radioactive decay from the last r-process event. This together with $t_{\rm int}$ gives the contributing fraction from  prior events to meteoritic $^{244}$Pu abundance.

$t_{\rm LE}$ is deduced from the comparison of the ratio of unstable to stable r-process isotopes between meteoritic abundances and the theoretical production yields. Since $t_{\rm LE}$ critically depends on the assumed production ratio, we need several combinations of unstable/stable isotopes to make an independent estimate of $t_{\rm LE}$.  For instance, the useful isotope combinations are $^{247}{\rm Cm}/^{235}{\rm U}$, $^{129}{\rm I}/^{127}{\rm I}$, and $^{244}{\rm Pu}/^{238}{\rm U}$ where the unstable isotopes have half-lives of $1.56\times 10^7$yr, $1.57\times 10^7$yr, and  $8.1\times 10^7$yr, respectively. From $^{244}{\rm Pu}/^{238}{\rm U}$ and $^{129}{\rm I}/^{127}{\rm I}$, a relatively long interval of $t_{\rm LE}\approx$ 100 Myr is obtained \citep{Reynolds_60, Wasserburg_69, Clayton_83, Dauphas_05}. More recently, using the latest nucleosynthesis results of $^{129}{\rm I}$ and $^{127}{\rm I}$, \citet{Lugaro_14} deduce $t_{\rm LE}=$109 Myr together with $t_{\rm LE}=$123 Myr from $^{247}{\rm Cm}/^{235}{\rm U}$. Though the value of $t_{\rm LE}$ thus obtained should be revised as will be discussed in \S 2.3, we here adopt $t_{\rm LE}=$120 Myr as an initial input. In addition, we assume $t_{\rm int}$=200 Myr, which will yield the adopted $t_{\rm LE}$ as one of the plausible cases.

From $t_{\rm LE}=$120 Myr, the remaining fraction of $^{244}{\rm Pu}$ from the last event at the ESS is estimated to be 35.8\%. Then, from both $t_{\rm LE}=$120 Myr and $t_{\rm int}$=200 Myr, we deduce that the fraction of $^{244}{\rm Pu}$ from the last event with respect to the total $^{244}{\rm Pu}$ in meteoritic abundance is 82\%, which is confirmed to be a large fraction as expected. These values combined with the prior estimate of $8.90\times 10^{-13}$ results in $2.03\times 10^{-12}$ as the mass fraction of $^{244}{\rm Pu}$ in the ISM yielded from an injection by a single r-process event.
 
\subsection{Total frequency of r-process events until solar system formation}

Next, we relate the above $^{244}{\rm Pu}$ mass fraction to that of stable Eu nuclei by introducing the theoretical production ratio between the two nuclei. Production of radioactive r-nuclides such as Th, U, and Pu is investigated by \citet{Goriely_01}, and according to their latest nucleosynthesis results \citep{Goriely_16}, the updated production ratio for $^{244}{\rm Pu}$/$^{238}{\rm U}$ is 0.33. This value is identical to other estimates by \citet{Cowan_87} (=0.4) and \citet{Eichler_15} (=0.33).

On the other hand, regarding the existing r-nuclides in meteorites (or solar photosphere), we anticipate that the production pattern follows the solar r-process pattern. This hypothesis is justified by the observation that Galactic metal-poor halo stars exhibit the universality of r-process abundance distribution for heavy r-nuclides, which matches the solar r-process pattern \citep[e.g.,][]{Montes_07, Sneden_08}.  Accordingly, the relative production ratio for $^{238}{\rm U}$/Eu is inferred from their solar r-process ratio corrected by the radioactive decay for $^{238}{\rm U}$ \citep[Eu: 0.0984, $^{238}$U: 0.018 (atoms/$10^6$ Si);][]{Lodders_09} as well as by a 3\% contribution to Eu from s-process. The mass ratio is found to be 0.2955. This together with $^{244}{\rm Pu}$/$^{238}{\rm U}$=0.33 results in $^{244}{\rm Pu}$/Eu=0.097. Then we finally obtain the value of Eu mass fraction due to enrichment by a single r-process event as $2.10\times10^{-11}$.

The solar r-process abundance of Eu, $3.75\times10^{-10}$, is the end result of cumulation of Eu by individual r-process events in the solar vicinity till the ESS. Thus, using the enrichment caused by each event, we can count the total number of events till the ESS. Dividing $3.75\times10^{-10}$ by $2.10\times10^{-11}$, we obtain the total number of r-process events till the ESS as $\sim$18. Here we ignore the mixing process of the ISM on the $\sim$100 Myr timescale, which may allow Eu atoms to move into and out of the ISM yielding the solar system over $\sim$7.5 Gyr. Simulations of r-process enrichment combined with hydrodynamics of the ISM in the Galaxy are indeed awaited.

\subsection{Last r-process event time}

Since the total number $N_r$ of r-process event determines the abundances of stable r-nuclei at the ESS, $t_{\rm LE}$ is largely influenced by $N_r$. This is understood from the abundance ratio $R_{\rm LE}$ of a radioactive isotope to a stable isotope after the last event inside meteorites, which is given by the following equation \citep{Lugaro_14}; 
\begin{equation}
R_{\rm LE} = R_{\rm yield}\times\frac{1}{N_r}\times(1+\frac{e^{-t_{\rm int}/\tau}}{1-e^{-t_{\rm int}/\tau}}) \ \ \ ,
\end{equation}
where $R_{\rm yield}$ is the production ratio of each single event and $\tau$ is the mean lifetime (=half-life/$\ln 2$) of  a radioactive isotope. Thus, $t_{\rm LE}$ should be updated with our result. To calculate $t_{\rm LE}$, we need to assume $t_{\rm int}$ for the period of \ltsim 1 Gyr  before the ESS in addition to $N_r$. Then incorporating $N_r$=18 with $t_{\rm int}$=200 Myr before the ESS into calculations, we deduce $t_{\rm LE}$=137, 139, and 145 Myr from $^{129}{\rm I}/^{127}{\rm I}$, $^{247}{\rm Cm}/^{235}{\rm U}$, and $^{244}{\rm Pu}/^{238}{\rm U}$, respectively. Accordingly we here adopt $t_{\rm LE}$=140 Myr. This revised $t_{\rm LE}$ modifies the prior $^{244}{\rm Pu}$ remaining fraction of the last event in meteorites from 0.358 to 0.302. Then the resultant mass fractions of $^{244}{\rm Pu}$ and Eu by one r-process event are slightly modified to $2.41\times 10^{-12}$ and $2.49\times10^{-11}$, respectively. These new values result in $N_r \sim$ 15. Hereafter we adopt $N_r$=15, though this value can be improved further as there are still uncertainties in $t_{\rm LE}$ and $t_{\rm int}$. A further update of $t_{\rm LE}$ will be done with a variable $t_{\rm int}$ in the next section.

\section{Incorporation with the local star formation history}

\begin{figure}[t]
\vspace{0.2cm}
\begin{center}
\includegraphics[width=8.5cm,clip=true]{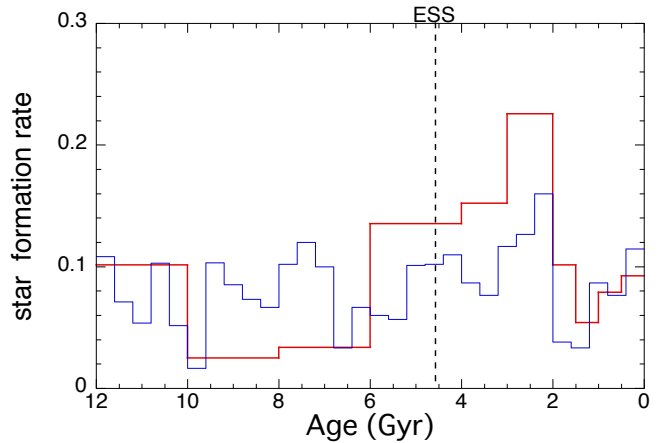}
\end{center}
\caption{Observed star formation rate (SFR) as a function of age in the solar vicinity. Red and blue lines are the ones derived by \citet{Cignoni_06} and \citet{Rocha_00}, respectively. For the result of \citet{Rocha_00}, the integrated SFR over $t$=12$-$15 Gyr is equally redistributed to individual bins for the time interval $t$=10$-$12 Gyr to achieve  compatibility between Galactic age and cosmic age. 
}
\end{figure}

Incorporating the total number of r-process events till the ESS into the observed local star formation (SF) history, we calculate the evolution of $^{244}{\rm Pu}$ density of the solar neighborhood over the entire period up to the present. Our approach has two steps: (i) we integrate the SFR from the beginning of the Galaxy to the time of ESS and bin by the inferred number of contributing r-process events during that time, and (ii) we integrate the SFR forward from the ESS time to the present, and at any time that the integrated number of newly formed stars reaches the threshold number, $^{244}$Pu is ejected into the ISM. We utilize the two observed SF histories obtained by \citet{Cignoni_06} and \citet{Rocha_00} because these two papers try to give the whole SF history while others limit it to the recent Gyrs. Their results are shown by red and blue lines, respectively, in Figure 1. Applying the above procedures to the SF histories, we find that each SF history results in 26 and 31 events of r-process production in total. Then from the assigned individual r-process events as a function of time, we calculate the time evolution of $^{244}{\rm Pu}$ density in the ISM (Fig.~2). Owing to the observed tendency that the SFR during the recent 2 Gyr is relatively low compared with at that over 2$-$5 (6) Gyr ago including the ESS epoch, $t_{\rm int}$ changes from $\sim 250$ Myr around at the ESS to $\sim 400$ Myr at the current time. A relatively lower SFR in recent times is also deduced from white dwarf cosmochronology \citep{Tremblay_14}. Since the oscillatory lowest level of $^{244}{\rm Pu}$ density is determined by the length of $t_{\rm int}$, the recent long $t_{\rm int}$ makes the $^{244}{\rm Pu}$ density drop to the value as low as the measurement in ocean archives \citep{Wallner_15} during the last 2 Gyr. The time interval of the latest r-process  event inferred from the $^{244}{\rm Pu}$ density measured in ocean sediments is 360 Myr. If we assume the SFR prior to the ESS time, a Poisson distribution of r-process events gives a no-event  probability of 0.2 within the time interval. On the other hand, the observed current low SFR lifts the probability to 0.4 and further to 0.65 for the upper limit of the sediment abundance.

\begin{figure}[t]
\vspace{0.2cm}
\begin{center}
\includegraphics[width=8.5cm,clip=true]{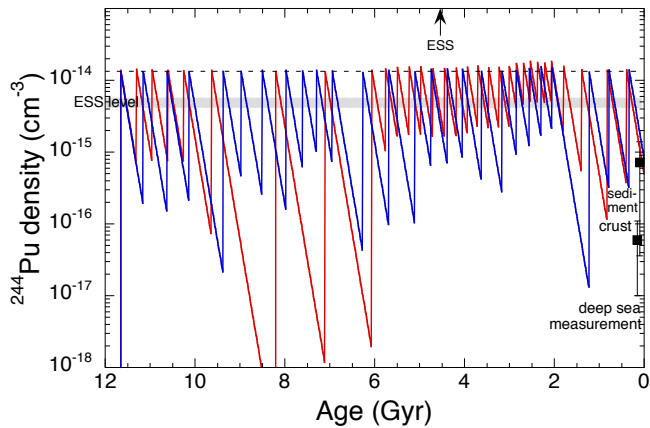}
\end{center}
\caption{Time evolution of $^{244}$Pu number density in the ISM of the solar vicinity. The different colored lines correspond to the results combined with the star formation history shown by the same color in Fig.~1. The current measurement values estimated from deep sea archives \citep{Wallner_15} are shown with error bars. The level of $^{244}$Pu density at the ESS is denoted by the shaded stripe. The initial $^{244}$Pu density by a single r-process event is also indicated.
}
\end{figure}

In addition, from our predicted $^{244}{\rm Pu}$ evolution, we deduce $t_{\rm LE}$ with a variable $t_{\rm int}$, which is calculated based on the SFR-$t$ relation in Fig.~1. Then, we eventually obtain $t_{\rm LE}$=130 Myr from the SF history  of \citet{Cignoni_06} and 140 Myr from that of \citet{Rocha_00}. This longer value of $t_{\rm LE}$ compared with the values thus far is mainly caused by the small $N_r$ till the ESS, which results in a high initial $^{244}$Pu/$^{238}$U ratio at the last r-process event owing to a low $^{238}$U abundance that reflects the cumulation of the past r-process events.

\section{Properties of r-process events}

\subsection{Propagation}

By comparing the deduced $^{244}$Pu density ejected in the ISM by a single r-process event with the $^{244}$Pu nucleosynthesis yield from a r-process site, we can estimate how much the ISM will be eventually mixed with the ejecta carrying r-process elements. Considering the implied rarity of frequency as a countable number of $\sim$30 events over the period of 12 Gyr in the solar vicinity together with a detailed discussion in a previous work \citep{Hotokezaka_15},  the r-process production site can be reasonably identified with an NS merger. Then, we assume a typical ejecta mass of an NS merger of 0.01 \ms and individual element production obeying the solar r-process pattern. In addition, we assume that light r-process elements such as Sr, Y, and Zr ($A\approx 90$) are not synthesized in an NS merger from an implication of elemental feature among Galactic halo stars \citep[][see also Monte et al.~2007]{Tsujimoto_14b}. According to the calculations claiming the production of $A$\gtsim 130 by an NS merger \citep[e.g.,][]{Bauswein_13}, we set the lower boundary at $A$=127 since the discussion based on the last r-process event (see \S 2.1) suggests that the production site of iodine is the same as that of heavier r-nuclei such as Pu.

The above hypotheses together with the radio-decay correction give the $^{238}$U mass fraction inside the ejecta of 2.55$\times10^{-3}$\msp. Then from $^{244}{\rm Pu}$/$^{238}{\rm U}$=0.33, the mass fraction of $^{244}$Pu is estimated to be 8.42$\times10^{-4}$, which is equivalent to a mass of 8.42$\times10^{-6}$\ms in the ejecta with a mass of 0.01 \msp. Combining this ejected mass of $^{244}$Pu with its resultant density in the ISM, the mass of ISM mixed with the r-process ejecta is deduced to be 8.42$\times10^{-6}$\ms/$2.41\times 10^{-12}\approx3.5\times10^6$\msp. It turns out that the mass mixed with the ejecta of an NS merger that expands with a velocity of 10\% to 30\% of the speed of light is much larger than that swept-up by an SN of $5.1\times10^4$\ms \citep{Shigeyama_98}. The corresponding radius of the cylindrical disk is found to be 370 pc, assuming a local surface density of gas (HI + H$_2$) of 8 \ms pc$^{-2}$ \citep{Koda_16}. The timescale to pervade the ISM could be as short as on the order of Myr \citep{Tsujimoto_14a} and thus negligible.

\subsection{Frequency}

Utilizing the result on the propagation of r-process ejecta, we can deduce the event frequency of r-process production with respect to the occurrence number of CCSNe. The current time interval between r-process production events is found to be about 400 Myr. Thus, it turns out that the total number of CCSNe inside a cylindrical disk with a 370 pc radius for a period of 400 Myr eventually yields one r-process production event. Here we assess a  local CCSN rate from two approaches. For the Galaxy in whole, the current SFR and CCSN rate are estimated to be 1.65 \ms yr$^{-1}$ \citep{Licquia_15} and 2.3 SNe per century \citep{Li_11}, respectively; these two establish the correlation that an SFR of 1 \ms yr$^{-1}$ is equal to 1.4 SNe per century. Then, from the present-day local SFR of 0.48 -- 1.1 \ms Gyr$^{-1}$pc$^{-2}$ \citep{Fuchs_09, Tsujimoto_11}, we obtain the CCSN rate of one per 2.1-4.8 Myr within a 100 pc radius. On the other hand, the recent detection of $^{60}$Fe in deep-sea sediments reveals that two CCSNe occurred 1.5-3.2 Myr and 6.5-8.7 Myr ago \citep{Wallner_16} within a distance of 100 pc \citep{Breitschwerdt_16}. Then, putting the two results together, we finally adopt the CCSN rate of one per 4 Myr within a 100 pc radius.

The derived CCSN rate gives 1370 CCSNe in total inside a cylindrical disk with a 370 pc radius for a period of 400 Myr. This estimate is equivalent to the rate of r-process production by NS merger of one per 1370 CCSNe. The rate is in fairly good agreement with the estimate of one per 1000$-$2000 CCSNe made from the correlation of stellar abundance between [Fe/H] and [Eu/H] in dSphs as well as from Galactic chemical evolution \citep{Tsujimoto_14a}. We reanalyze its frequency from abundance correlation between [Mg/H] and [Eu/H] among Galactic stars to avoid uncertainties in the chemical enrichment process in dSphs and contribution from SNe Ia. By inputting Mg yield from a CCSN deduced by the observed two quantities, i.e., an average Fe mass of 0.07 \ms \citep[light curve analysis:][]{Hamuy_03} and the plateau of [Mg/Fe]=0.4 among halo stars, we obtain the frequency of one per $\sim$1400  CCSNe for Eu production.

\acknowledgements

This work benefited from support by the National Science Foundation under Grant No. PHY-1430152 (JINA Center for the Evolution of the Elements), and was supported by JSPS KAKENHI Grant Number 15K05033, 15K13602, and 16H04081.

\end{document}